\documentclass[sigconf,nonacm]{acmart}

\usepackage{tikz}

\AtBeginDocument{%
  \providecommand\BibTeX{{%
    \normalfont B\kern-0.5em{\scshape i\kern-0.25em b}\kern-0.8em\TeX}}}

\makeatletter
\def\@ACM@checkaffil{
    \if@ACM@instpresent\else
    \ClassWarningNoLine{\@classname}{No institution present for an affiliation}%
    \fi
    \if@ACM@citypresent\else
    \ClassWarningNoLine{\@classname}{No city present for an affiliation}%
    \fi
    \if@ACM@countrypresent\else
        \ClassWarningNoLine{\@classname}{No country present for an affiliation}%
    \fi
}
\makeatother

\begin{document}

\title{S3C2 SICP Summit 2025-06: \\ Vulnerability Response Summit}

\author{Anna Lena Rotthaler$^{**}$, Simon Oberthür$^{\ddagger\ddagger}$, Juraj Somorovsky$^{\ddagger\ddagger}$, Kirsten Thommes$^{\ddagger\ddagger}$, Simon Trang$^{\ddagger\ddagger}$, Yasemin Acar$^{\dagger \& \ddagger\ddagger}$, Michel Cukier$^{\ddagger}$, William Enck$^{*}$, Alexandros Kapravelos$^{*}$, Christian Kästner$^{\mathsection}$, Dominik Wermke$^{*}$, Laurie Williams$^{*}$}

\def \authors{Anna Lena Rotthaler, Simon Oberthür, Juraj Somorovsky, Kirsten Thommes, Simon Trang, Yasemin Acar, Michel Cukier, William Enck, Alexandros Kapravelos, Christian Kästner, Dominik Wermke, Laurie Williams}

\affiliation{
    \institution{ $^*$North Carolina State University, Raleigh, NC, USA}
}
\affiliation{
    \institution{$^\dagger$George Washington University, DC, USA}
}
\affiliation{
    \institution{$^{\ddagger\ddagger}$Software Innovation Campus Paderborn, Paderborn University, Paderborn, Germany}    
}

\affiliation{
    \institution{$^{**}$Paderborn University, Paderborn, Germany}    
}

\affiliation{
    \institution{$^\ddagger$University of Maryland, College Park, MD, USA}
}
\affiliation{
    \institution{ $^\mathsection$Carnegie Mellon University, Pittsburgh, PA, USA}
}

\renewcommand{\shortauthors}{Software Innovation Campus Paderborn and Secure Software Supply Chain Center (S3C2)}
\renewcommand{\shorttitle}{S3C2 SICP Summit 2025-06: Industry Vulnerability Reponse Summit}

\begin{abstract}
  Recent years have shown increased cyber attacks targeting less secure elements in the software supply chain and causing significant damage to businesses and organizations. The US and EU governments and industry are equally interested in enhancing software security, including supply chain and vulnerability response. On June 26, 2025, researchers from the NSF-supported Secure Software Supply Chain Center (S3C2) and the Software Innovation Campus Paderborn (SICP) conducted a Vulnerability Response Summit with a diverse set of 9 practitioners from 9 companies. The goal of the Summit is to enable sharing between industry practitioners having practical experiences and challenges with software supply chain security, including vulnerability response, and helping to form new collaborations.
  We conducted five panel discussions based on open-ended questions regarding 
  experiences with vulnerability reports, tools used for vulnerability discovery and management, organizational structures to report vulnerability response and management, preparedness and implementations for Cyber Resilience Act\footnote{\url{https://digital-strategy.ec.europa.eu/en/policies/cyber-resilience-act}} (CRA) and NIS2\footnote{\url{https://digital-strategy.ec.europa.eu/en/policies/nis2-directive}}, and bug bounties.
  The open discussions enabled mutual sharing and shed light on common challenges that industry practitioners with practical experience face when securing their software supply chain, including vulnerability response. In this paper, we provide a summary of the Summit. Full panel questions can be found in the appendix.
\end{abstract}


\maketitle

\begin{tikzpicture}[overlay, remember picture]
\node[anchor=north west, 
      xshift=17.5cm, 
      yshift=-2.1cm] 
     at (current page.north west) 
     {\includegraphics[width=2.1cm]{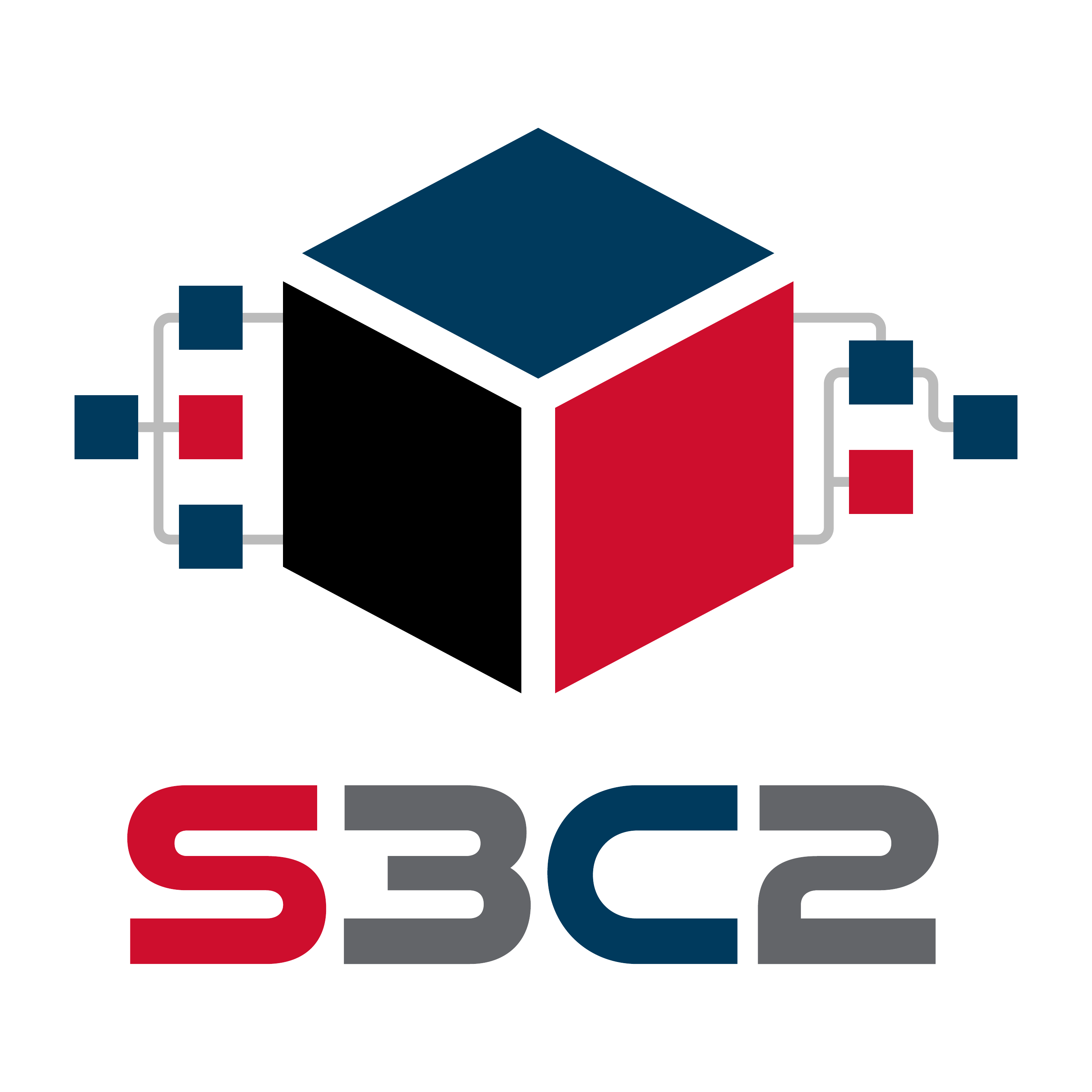}}; 
\end{tikzpicture}

\begin{tikzpicture}[overlay, remember picture]
\node[anchor=north east, 
      xshift=-17.5cm, 
      yshift=-3.1cm] 
     at (current page.north east) 
     {\includegraphics[width=2.1cm]{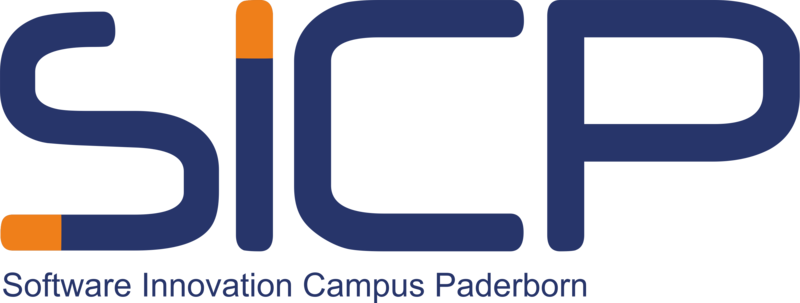}}; 
\end{tikzpicture}

\section{Introduction}

  Recent years have shown increased cyber attacks targeting less secure elements in the software supply chain and causing significant damage to businesses and organizations. Past well-known examples of software supply chain attacks are the SolarWinds or log4j incidents that have affected thousands of customers and businesses. On June 26, 2025, Software Innovation Campus Paderborn, together with NSF-supported Secure Software Supply Chain Center (S3C2)\footnote{\url{https://s3c2.org}} researchers, conducted a one-day Vulnerability Response Summit in Germany with 9 practitioners from 9 companies.  The goal of the Summit is to enable sharing between industry practitioners having practical experiences and challenges with software supply chain security, including vulnerability response; to help form new collaborations between industrial organizations and researchers; and to identify research opportunities.

  Summit participants were recruited from 9 companies, intentionally in diverse domains and having various company maturity levels and sizes.  Attendance is limited to one participant per company to keep the event small enough that honest communication between participants can flow. The Summit was conducted under the Chatham House Rules, which state that all participants are free to use the information discussed, but neither the identity nor the affiliation of the speaker(s), nor any other participant may be revealed. As such, none of the participating companies are identified in this paper.  

  The Summit consisted of one keynote presentation and five panels. Before the Summit, participants completed a survey to vote on the topics of the five panels.  As such, the panel topics represent the challenges faced by practitioners. Based upon personal preferences expressed in the survey, three to four participants were selected to begin each 45-minute panel discussion with a 3-5 minute statement.  The remaining minutes of each panel were spent openly discussing the topic. The questions posed to the panelists appear in the Appendix.   

 Researchers (four professors, one scientific employee, one Ph.D. student) took notes on the discussions. We jointly created a first draft summary of the discussion based on these notes. The draft was reviewed by all attendees and authors, including all S3C2 researchers, who are experts in software supply chain security.

The next five sections provide a summary of the Vulnerability Response Summit.

\section{Experiences with Vulnerability Reporting}
Attendees reported a change over time in companies publishing vulnerabilities: five years ago, companies were hesitant to publish vulnerabilities for the fear of being perceived as providing low-quality products; this recently changed towards transparency and open communication to increase customer trust.

Most attendees mentioned that they use a single (support) email address that can be used to report vulnerabilities, and recommended this strategy so that the barrier to finding where to report vulnerabilities is very low.

A challenge mentioned was informing customers of vulnerabilities, especially in B2C companies, because they may not know the end users of their products. The strategy overall was ``inform the customers, if you know them, if not, inform all'', i.e., publish the vulnerability. 

For many attendees, vulnerabilities are reported by customers, not security researchers, hackers, or bug bounty hunters.

\subsection{Assessing Vulnerabilities}
Attendees agreed that the biggest challenge when receiving vulnerabilities is to assess the severity of the vulnerability: automation is often not possible, and the decision-maker needs to know how the software works and also how and in which context it is used by specific customers. Attendees discussed a lack of talent that bridges deep product knowledge and security knowledge.  The panel discussed different strategies to deal with this challenge: 
Criticality decisions are often made using established methods, e.g., the CVSS score.
Criticality decisions are complex because you need to know the context for all product use. The problem is that employees need both security and software use knowledge to be able to make this assessment. Employees are likely to escalate to ``more severe'' because they need trust / confidence to downgrade it. Participants mentioned assessment strategies such as call graph analysis to see if the vulnerable code is reachable / if the parameter associated with the vulnerability can be changed externally.

\subsection{Fixing Vulnerabilities}
Attendees mention that the time to fix is typically given by the reporter of the vulnerability, typically 90 days. Attendees consider vulnerabilities ``end to end'', from report until publication. Depending on the communication with stakeholders and suppliers, 90 days can be appropriate or short. One attendee reported on an internal experiment to have a faster timeline and figure out if it is possible for them to optimize.

Attendees mentioned the CRA as a challenge because they assume that ``one active vulnerability per team'' may go up to 20. They are interested in whether the standard time to fix will stay at 90 days and assume that it may become longer once companies have published their contact data and receive a large volume of vulnerability report information (as well as unrelated information).

Attendees mentioned handling different levels of criticality differently: sometimes, for high criticality, they help customers to mitigate even before there is a patch; the criticality score determines the time to remediate. This may also be different per product; some products need patching, others may have the vulnerability, but it cannot be exploited, so the patch may be delayed.

\subsection{Open Questions}
At the end of the panel, some open questions and problems remained:

\begin{itemize}
    \item How to deal with the differences in B2C and B2B contexts?
    \item How can we prioritize vulnerabilities and find employees who can rate/prioritize (who have security knowledge and knowledge about product use)?
    \item How can we improve the complex manual process of vulnerability assessment?
    \item What are standards for reporting? How can we work towards effective standardization?
\end{itemize}

\section{Tools}
Attendees reported mixed feelings about tools for vulnerability management. Some attendees mentioned using tools that check vulnerability databases, which only work for known vulnerabilities. They are worried that the US will stop maintaining those databases (e.g., the CVE database is maintained by the MITRE Corporation, which is funded by the US government).
The panel discussed the use of tools critically: ``many tools deliver many outputs'', and the question remained how to handle this output. One solution mentioned was to reduce the tools to reduce the output that needs to be handled. One attendee mentioned that there are internal AI tools to check code security, other attendees mentioned using specific software composition analysis tools, and another attendee mentioned only searching components for vulnerabilities, but not their own code. Most of the tools mentioned are ``after build'' and only assess security after the development phase.

\subsection{Choosing Tools}
Attendees reported that different teams use different tools. They are chosen by first collecting challenges and needs, then evaluating the tools and selecting those that can deliver. Tool usability is mentioned as very critical: false positives can create overwhelm. If specific scanners have a high number of false positives, they are unusable. One attendee reported that at the moment, commercial tools are cheaper for them than open source tools because they do not need to train/update/maintain them. Another usability problem mentioned was that tools are often not compatible because they work on source code, but the company needs to scan binaries. Another challenge mentioned is that different tools provide different results and there are no reporting standards. Attendees would like to have a single source of truth, ideally achieved by reporting standards like SPDX/CycloneDx and VEX, which could be imported and exported to have data exchange between tools.

\subsection{Open Questions}
At the end of the panel, some open questions remained:

\begin{itemize}
    \item How can we improve the usability of tools?
    \item How to prioritize vulnerabilities?
    \item How can we get a ``single source of truth''? How can we work towards the missing standard interpretation for communication between tools?
    \item How can we get an ``SBOM with different security zones''? How can SBOMs capture ``this cannot be reached, it is encapsulated in the backend''?
    \item How can we reduce false positives?
\end{itemize}

\section{Organization of Vulnerability Reporting}
Most attendees mentioned having a dedicated structure for vulnerability reporting. One attendee mentioned building it in the context of ISO 27001. They reported a similar structure, with one central email address, dedicated roles such as CISO, and a predefined escalation structure. Internally, attendees mentioned that every employee knows where and how to report without any negative consequences (they do not personally have to fix it). 

\subsection{Reaction}
Attendees report that the reaction to a vulnerability report depends on the severity of the vulnerability. Worst case is isolation (take the system offline). The usual case is that the vendor has a fix, and if the vulnerability is critical, they will patch in production with the risk that the patch causes problems, or wait if there is time to test. Attendees report that low-level vulnerabilities are daily business, and that major incidents will be assessed individually.

The panel discussed the challenge of estimating risk/damage. They mentioned the following parameters: the number of affected devices, danger to life, criticality based on CVSS scores, based on experience, think in categories. Attendees agreed that it is impossible to rely on automated tools for risk assessment because the tools can produce false positives, do not know the context, and rely on heuristics. Therefore, the main problem mentioned is the human effort needed for pre-assessment, finding employees with the know-how/expertise, and the number of incidents that have to be handled by humans.

\subsection{Open Questions}
At the end of the panel, some open questions remained:

\begin{itemize}
    \item How can we incentivize employees to report vulnerabilities without burdening them (with unwillingly having to fix them)?
    \item How can we aggregate / automatically pre-assess reported vulnerabilities?
    \item How can we process information? What is the risk? Societal damage?
    \item How can we get from manual and ``triage'' risk assessment to automated risk assessment?
    \item How can we measure ``Was it really critical? Did we pay too much to fix too quickly?''
\end{itemize}

\section{Cyber Resilience Act (CRA)/NIS2}
The Cyber Resilience Act (CRA)~\cite{CRA} introduces regulations and standards for software and hardware for the European market. The NIS2 Directive is a continuation and expansion of the previous EU cybersecurity directive, NIS~\cite{NIS2}.
Attendees clearly separated between CRA and NIS2.

\subsection{NIS2}
For NIS2, some attendees ``feel'' well prepared but mentioned that information is coming too late and is too imprecise. Other attendees who are ISO 27001 compliant feel well-prepared and on a good track. Standards that are common in banking/insurance and similar industries were already demanded by the customers. Impacts will be visible in other sectors where security standards were not already enforced.
Attendees mentioned that NIS2 is already applied in other countries, so that companies can implement anticipated big changes now, and small changes maybe later. The BSI also offers events on news. 

\subsection{CRA}
Attendees expressed uncertainty regarding the CRA. Some of the attendees reported uncertainty because they do not yet know whether they will be affected and which class they are in (important or critical products). There are currently no consultants on this topic because it is too fresh. Attendees observe the situation and are preparing to read a significant volume of documents once the CRA is published, and probably hire lawyers and/or consultants. Thus, the CRA will be very expensive for companies: smaller companies may not be able to comply without making parts more expensive. 
The CRA will also have an impact on responsibility: Attendees point out that the CRA says that if your product has a vulnerable component, you are responsible, and are wondering how to organize this. They discussed that it is very unclear who in the middle of the supply chain takes the risk. They assume that global players will push this down, while small enterprises will be unable to push the risk up, leaving medium-sized enterprises to shoulder the risk. Post-release security becomes more important until decommissioning.
One attendee suggests that organizations focus and center on core products to serve only specific branches and be compliant.

Attendees are generally not worried about the obligation to report vulnerabilities within 24 hours via the ENISA platform because all the processes are in place. They take into consideration, though, that if there is no time to analyze appropriately, ENISA will be flooded with low-quality reports.

Attendees reported buying trainings, e.g., from Fraunhofer, and establishing roles/experts. Roles and processes according to IEC 62443-4-1 and security champions were mentioned. The panel agreed that security needs experts and that companies need to be enabled.

\subsection{Open Questions}
At the end of the panel, some open questions remained:

\begin{itemize}
    \item How to handle the overwhelming volume of new guidance for CRA while consulting around this does not yet exist? (Harmonizing standards are in development)
    \item How can existing standards and CRA be harmonized? 
    \item How can we handle problems with time, information, interpretation, and processing until deadlines, especially in light of missing ``legal'' information?
\end{itemize}

\section{Bug Bounties}
From the security researcher's perspective, finding contact data is still one big problem when reporting vulnerabilities (though they acknowledge that it is better than before), and another big problem is that they have mixed feelings when reporting. In their opinion, bug bounties \textit{could} help, but are not currently very useful (for researchers). From the company's perspective, bug bounty programs commercialize vulnerability search, paying the vulnerability reporter for their work. Most attendees did not participate in a bug bounty program; they were confronted with bug bounty providers that want the company to opt in, or confronted with people asking ``Do you have a bug bounty? If not, it's not necessarily worth it for me to look at your product''. For most attendees, the value of bug bounty programs is not currently apparent. 

\subsection{Bug Bounty Programs vs Pentesting}
The panel discussed the costs and benefits of bug bounty programs as well as the value of bug bounty programs in contrast to pentesting. Attendees reported mixed feelings but generally agreed that pentesting is preferred with less maturity, and bug bounty programs are only possible with very high maturity. For B2B contexts, the products may be inaccessible to external bug hunters. Attendees mentioned that bug bounties may demonstrate high maturity and trustworthiness to customers if the targeted customers interpret the existence of a bug bounty program correctly (somewhat high security literacy required).
The panel discussed bug bounty programs in the context of the CRA: they are recognized by the regulatory authorities if they are part of a systematic, coordinated process for vulnerability handling (e.g., CVD, PSIRT).

\subsection{Open Questions}
At the end of the panel, some open questions remained:

\begin{itemize}
    \item How do bug bounties work for B2B?
    \item Is the presence of bug bounty programs effective marketing?
\end{itemize}

\section{Executive Summary}
\label{executive}
Managing and remediating vulnerabilities is important, timely, and targeted by current regulations. Attendees are preparing for the CRA and expecting that it will increase the baseline for security. They are confident in their preparation for NIS2. 
The major complication in vulnerability management and mitigation was discussed as dealing with a large volume of information, and needing specialized security knowledge, as well as in-depth product knowledge to correctly assess and prioritize handling vulnerabilities. Attendees expect that this challenge will increase with the CRA. They are not currently perceiving AI as helpful in this challenge. SBOMs and SCAs that provide information on vulnerable components in the build process were discussed as helpful.

\begin{acks}
A big thank you to all Summit participants. We are very grateful for being able to hear about your valuable experiences and suggestions. The Summit was organized at Software Innovation Campus Paderborn by Yasemin Acar, Simon Oberthür, Juraj Somorovsky, Kirsten Thommes, Simon Trang, and Anna Lena Rotthaler, modeled on previous summits organized by Laurie Williams, and was recorded by Anna Lena Rotthaler, Juraj Somorovsky, Simon Oberthür, and Yasemin Acar. The summit was sponsored by Software Innovation Campus Paderborn. This material is based upon work supported by the National Science Foundation Grant Nos. 2207008, 2206859, 2206865, and 2206921.
These grants support the Secure Software Supply Chain Summit (S3C2), consisting of researchers at North Carolina State University, Carnegie Mellon University, University of Maryland, George Washington University, and Paderborn University. Any opinions expressed in this material are those of the author(s) and do not necessarily reflect the views of the National Science Foundation nor Software Innovation Campus Paderborn.
\end{acks}

\bibliographystyle{ACM-Reference-Format}
\bibliography{literature}

@article{NIS2,
title = {What is NIS2?},
author={NIS2 Directive},
journal={https://nis2directive.eu/what-is-nis2/},
year={2025}
}

@book{CRA,
author = {European Commission and Directorate-General for Communications Networks, Content and Technology},
title = {Cyber resilience act – New EU cybersecurity rules ensure more secure hardware and software products},
publisher = {European Commission},
year = {2022},
doi = {doi/10.2759/543836}}

\appendix

\section{Full Survey Questions for Panel}
\label{questions}
\begin{enumerate}
\item Do you have experience with vulnerability reporting (from external sources)? What happened? What went well, what did not go well? What could be done better, what have you changed since then? How long does it take to fix the vulnerabilities? What does that depend on? How does coordination with customers and dependencies work? Do you have experience with the BSI's Coordinated Vulnerability Disclosure Guideline?

\item What tools do you use to manage vulnerabilities automatically? How are they integrated into your processes? How do you select them? How satisfied are you with them? How helpful are the tools in detecting, prioritizing and repairing? How do security updates, patch management, and distribution work?

\item How are you organized when it comes to vulnerability reporting? What is procedurally anchored / built into the governance framework? How could you be contacted if someone finds vulnerabilities in your products or infrastructure? Who would be contacted (which role)? How is the response organized internally, what are the internal responsibilities and resources? Do you have PSIRT/CSIRT teams? How does the risk assessment work?

\item How prepared are you for the Cyber Resilience Act? And for other directives such as NIS2? How do you handle legal risks when dealing with vulnerabilities? How is your corporate culture preparing for new, short-term reporting obligations (within 24 hours to ENISA)? Are there uncertainties in the interpretation of the changing legal situation? How do you generally organize awareness and training on the topic of vulnerability response?

\item Do you have any experience with bug bounties? How did you organize them, what did you choose, do you use certain platforms? What is your experience with the quality, relevance, and scope of the bug reports?
\end{enumerate}

\end{document}